\documentclass[
]{ceurart}

\usepackage{listings}
\begin{document}


\conference{In: 
B. Combemale, G. Mussbacher, S. Betz, A. Friday, I. Hadar, J. Sallou, I. Groher, H. Muccini,
O. Le Meur, C. Herglotz, E. Eriksson, B. Penzenstadler, AK. Peters, C. C. Venters. 
Joint Proceedings of ICT4S 2023 Doctoral Symposium, Demonstrations \& Posters Track and Workshops. Co-located with ICT4S 2023. Rennes, France, June 05-09, 2023.}
\copyrightyear{2023}
\copyrightclause{Copyright for this paper by its authors.
  Use permitted under Creative Commons License Attribution 4.0
  International (CC BY 4.0).}

\title{Towards a methodology to consider the environmental impacts of digital agriculture.}

\author[1]{Pierre {La Rocca}}[%
email=pierre.la-rocca@u-bordeaux.fr,
]
\cormark[1]
\address[1]{Univ. Bordeaux, CNRS, Bordeaux INP, INRIA, LaBRI, UMR 5800, F-33400 Talence, France}

\cortext[1]{Corresponding author.}

\begin{abstract}
Agriculture affects global warming, while its yields are threatened by it. Information and communication technology (ICT) is often considered as a potential lever to mitigate this tension, through monitoring and process optimization. However, while agricultural ICT is actively promoted, its environmental impact appears to be overlooked. Possible rebound effects could put at stake its net expected benefits and hamper agriculture sustainability. By adapting environmental footprint assessment methods to digital agriculture context, this research aims at defining a methodology taking into account the environmental footprint of agricultural ICT systems and their required infrastructures. The expected contribution is to propose present and prospective models based on possible digitalization scenarios, in order to assess effects and consequences of different technological paths on agriculture sustainability, sufficiency and resilience. The final results could be useful to enlighten societal debates and political decisions.
\end{abstract}


\maketitle

\section{Research project}
\subsection{Research context}

Agriculture affects global warming through the greenhouse gas (GHG) it emits \cite{european_environment_agency_eea_2022}. At the same time, its yields are affected by it according to \cite{ray_climate_2019}. In this context, information and communication technology (ICT) is seen by a part of political institutions \cite{muench_towards_2022}, agro-industries and academics as a potential lever to make agriculture more efficient, more resilient and more sustainable. The digitalization of agricultural processes is called \textit{digital agriculture}. The ICT used in this specific context is designed as \textit{Smart Farming Technologies} (SFT). Often cited SFT are robots, Internet of Things (IoT) \cite{farooq_role_2020} and Artificial Intelligence (AI).
Expected benefits from digital agriculture are summarized by \cite{moysiadis_smart_2021} as the increase in production, the decrease in cost by reducing the inputs needed such as fuel, fertilizer and pesticides, the reduction in labour efforts, and finally improvement in the quality of the final products. In other words, digital agriculture is expected to bring sustainability by process optimization, monitoring and traceability.
While SFT are being increasingly adopted by farmers in North America and Europe \cite{nowak_precision_2021}, their benefits to produce a fairer and more sustainable food are still debated. According to \cite{kernecker_experience_2020}, SFT lack concrete examples proving their added value to farmers. In the ideas of \cite{schnebelin_how_2021} and \cite{bellon_maurel_agriculture_2022}, the context of production need to be taken into account and ICT itself cannot be a sufficient lever if not used in more sustainable means of production.
Hence, SFT benefits could depend on the production context where they are used, the functions that they need to address, and the technological paradigm used to address those functions. 
The authors of \cite{hilbeck_aligning_2022} argues that digital solutions need to be used in agroecological context and centred on farmers' needs to be a potential sustainability lever. This last point is also supported by the authors of \cite{davies_co-creating_2014}, that add that developing ICT solutions in isolation from agricultural realities run the risk of hampering rather than advancing possibilities for sustainability transitions in the food system.

If digital agriculture might be seen as a sustainable solution, the environmental footprint linked to ICT seems rarely considered by itself. Yet, ICT device manufacturing has a significant environmental footprint \cite{freitag_real_2021}, and a global adoption of digital agriculture could lead to a massive deployment of complex electronic devices in the environment. This situation would not be sustainable following the conclusions of \cite{pirson_assessing_2021} regarding IoT, as it could cause digital rebounds \cite{coroama_digital_2019} at a scale cancelling the expected benefits of digital agriculture.

\subsection{Research objectives}
In this context, by considering the peculiar environmental footprint of the ICT needed by digital agriculture and the possible rebound effects it could generate, it seems relevant to ask what kind of ICT, deployed to which scales, and for which agricultural systems could bring a certain sustainability, without threatening the sufficiency and the resilience also required by agricultural systems.

The main goal of this research is to set up of a methodology taking into account the environmental footprint of agricultural ICT systems, and their consequences on a systemic scale. This methodology should help us to develop present and prospective models and scenarios, in order to evaluate and compare possible technological paths. The obtained results are expected to enlighten societal debates and political decisions.  

Currently at its earlier stage, this research in Computer Science is conducted at the Laboratoire Bordelais de Recherche en Informatique (LaBRI), under the supervision of Aurélie Bugeau (LaBRI), Gaël Guennebaud (Inria) and Anne-Laure Ligozat (LISN). It is realized in the Image and Sound department, because of the current disciplinary fields of its supervision, but also because of the lack of an ICT sustainability department.

\section{Methods and first results}
\subsection{Research Approach}

Our general methodology is inspired by the works of \cite{coroama_methodology_2020} and \cite{bergmark_methodology_2020}.
It is adapted to the farm complexity, where several companies can propose several ICT services working on a same territory. This while calling for different infrastructures not always interoperable. 
The goal of this methodology is to lead to the creation of parametric models able to inform decisions on the adoption of digital agriculture paths. Those models aim to go beyond individual life cycle assessment (LCA) studies to provide more systemic visions at different territorial scales. To adopt a more systemic approach implies broadening the environmental impact spectrum by looking at more indicators than carbon emissions and energy consumption indicators. Relevant additional indicators could be water consumption or potential metallic depletion linked to ICT infrastructures. Such models could help characterize absolute environmental impacts of digital agriculture systems and few associated indirect effects.
The main steps of the proposed methodology are the definition of a baseline and its boundaries, the identification of relevant case studies and the effect assessment of possible prospective trends.
Whether the digitization of agriculture being a matter of growing interest, this trend is far from being new according to \cite{farooq_role_2020}. Hence, the definition of a baseline need to take into account the already existing digitization. We then use as a general baseline the average ICT used on French farm in 2025. This baseline will further be refined according to considered case studies. After a first environmental footprint assessment, this baseline will be useful to compare different technological scenarios at the 2035 horizon. Inventory studies, as the one proposed by \cite{schnebelin_usages_2022} as well as field surveys will help us to acquire associated general data. 

The boundaries definition enables us to define the ICT infrastructures taken into account, the territory and the agricultural systems and steps where they are being deployed as well as the temporal perspective where they can evolve. Because different ICT systems can be used to address a same issue, we adopt a functionalist approach to compare different ICT systems addressing a same function. Approaching digital ICT by functions enables us to consider complex systems of embedded different technologies interacting to solve a same function instead of isolated technologies. This complex approach seems to better reflect systems like those presented by the works of \cite{navarro_systematic_2020} and \cite{pretto_building_2021}. Looking at the ICT infrastructures required by a system to realize a function firstly gives us the possibility to assess its environmental impact magnitudes. It also gives us a way to consider the different functions a same ICT infrastructure is expected to address.


\subsection{First Results}
The thesis started in October 2022. A first work consisted in a global synthesis to discover the multiplicity behind digital agriculture systems.   
A first use case dedicated to cattle individual identification and health metrics systems is currently realized to build and test our general methodology.
We also conducted field surveys on a farm and at agricultural events to characterize current and expected trends of agricultural ICT.

\subsection{Future works}

In the near future, we aim to propose a first model comparing and RFID plus IoT system with one using AI and computer vision. This model will address two complementary functions regarding AI infrastructure possibilities, which are individual identification and health metrics. To do so, our preliminary work needs to be enriched with IoT collars. Then, we would like to consider culture spraying in order to test the conceptualized method on another use case. 

Environmental metrics linked to software processes will need to be considered. If current systems mainly focus on edge computing, our field studies showed that a future trend for digital agriculture companies is to externalize computing to owned cloud platforms. 

In the long time, we plan to compare our 2025 baselines to different 2035 scenarios, proposing different technical systems scales. Those scenarios will be inspired by the European commission plans \cite{muench_towards_2022} or the projections of \cite{agrotic_5g_2021}.

\section*{Conclusion}

Our research aims to propose methods and models to better assess the environmental footprint of digital agriculture systems. Filling this current gap directly contributes to a better understanding of how ICT can help to mitigate the uncertainties between climate change and agriculture. To do so, we will rely on and adapt to the context of agriculture methods used in ICT environmental effects assessments. Participating to ICT4S doctoral symposium is an opportunity to present our research to people concerned by the role that can play ICT in sustainability. Presenting our work there would be the occasion for us to formalize them toward an initiated audience, and to get from it external yet relevant feedback. 

\bibliography{references}

\end{document}